\documentclass[prb,twocolumn,showpacs,preprintmunbers,amsmath,amssymb]{revtex4}
\usepackage{graphicx}
\usepackage{dcolumn}
\usepackage{bm}
\begin{document}
\title{Neutron Scattering Study of \\
Relaxor Ferroelectric (1-$x$)Pb(Zn$_{1/3}$Nb$_{2/3}$)O$_{3}$-$x$PbTiO$_{3}$}

\author{D. La-Orauttapong}
\author{J. Toulouse}
\affiliation{Department of Physics, Lehigh University, Bethlehem, Pennsylvania 18015-3182}

\author{Z.-G. Ye}
\author{W. Chen}
\affiliation{Department of Chemistry, Simon Fraser University, Burnaby, British Columbia,
Canada V5A 1S6}

\author{R. Erwin}
\affiliation{NIST Center for Neutron Research, NIST, Gaithersburg, Maryland 20899-8562}

\author{J.L. Robertson}
\affiliation{Oak Ridge National Laboratory, Solid State Division, Oak Ridge, Tennessee
37831-6393}

\begin{abstract}
Neutron \textit{elastic} diffuse scattering experiments performed on Pb(Zn$_{1/3}$Nb$_{2/3}$)O$_{3}$ (PZN) 
and on its solid solution with PbTiO$_{3}$(PT), known as PZN-$x$PT, with $x$=4.5\% and 9\% around many reflections 
show that diffuse scattering is observed around reflections with mixed indices in the transverse and diagonal 
directions only. From the width of the diffuse scattering peak a correlation length is extracted. In PZN, we have 
reported that the diffuse scattering is more extended in the transverse than in the diagonal 
directions~\cite{La-Orauttapong-etal:2001}. In the present work, the results show that the addition of PT leads 
to a broadening of the diffuse scattering along the diagonal, relative to the transverse directions, indicating 
a change in the orientation of the polar regions. Also, with the addition of PT, the polar nanoregions condense at 
a higher temperature above the transition than in pure PZN ($>$ 40~K), due to stronger correlations between 
the polar regions. Neutron \textit{inelastic} scattering measurements have also been performed on PZN-$x$PT. 
The results show the broadening of the transverse acoustic (TA) phonon mode at a small momentum transfer 
$q$ upon cooling. We attribute this broadening to the appearance of the polar nanoregions.
\end{abstract}

\pacs{77.84.Dy, 61.12.-q, 64.70.Kb, 77.80.Bh}

\maketitle

\section{Introduction}
Lead-based complex perovskite relaxor ferroelectrics, such as the mixed compound 
(1-$x$)Pb(Zn$_{1/3}$Nb$_{2/3}$)O$_{3}$-$x$PbTiO$_{3}$ (PZN-$x$PT)
and (1-$x$)Pb(Mg$_{1/3}$Nb$_{2/3}$)O$_{3}$-$x$PbTiO$_{3}$ (PMN-$x$PT), have recently attracted a great 
deal of attention because of their exceptional piezoelectric and dielectric properties. Particularly important 
are the compositions near the morphotropic phase boundary (MPB), where these properties are further enhanced. 
The high values of the piezoelectric and electrostrictive coefficients of PZN-$x$PT and PMN-$x$PT crystals of 
the compositions near the MPB, when measured in the proper orientations, are an order of magnitude greater 
than those of PbZr$_{1-x}$Ti$_{x}$O$_{3}$ (PZT) ceramics~\cite{Park-etal:1997,Viehland-etal:2001}. 
As a result, PZN-$x$PT and its magnesium analogue PMN-$x$PT are now being considered as the most promising 
candidates for the next generation of electromechanical transducers.

\begin{figure}[tbp]
\includegraphics[width=0.75\linewidth]{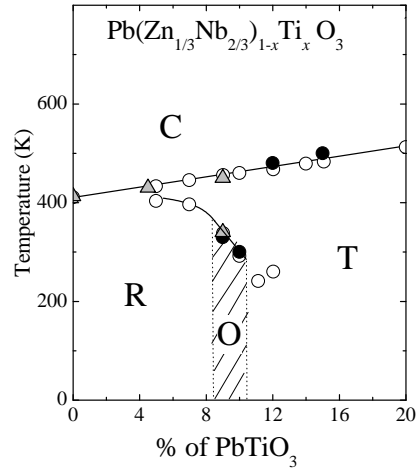}
\caption{Phase diagram of PZN-$x$PT around its MPB from Ref.~\onlinecite{La-Orauttapong-etal:2002}. 
The result of this work are also plotted as triangles showing the phase transition.}
\label{fig1}
\end{figure}

In the phase diagram of these systems, the MPB was initially believed to be an almost vertical boundary that 
separates the rhombohedral (space group $R3m$) from the tetragonal ($P4mm$) phase and located close to 
the value of $x \sim 10\%$ for PZN-$x$PT and $\sim 35\%$ for PMN-$x$PT~\cite{Kuwata-etal:1981,Shrout-etal:1990}. 
When crossing this boundary along the horizontal $x$ concentration axis, the system would change abruptly from one 
phase to the other. However, near the boundary, there is a delicate microstructural equilibrium that should favor easy local 
atomic rearrangements. This explains the very large coefficients observed. In addition to the known rhombohedral and 
tetragonal phases, a recent breakthrough has been achieved with the discovery of a sliver of a new monoclinic phase 
in PZT~\cite{Noheda-etal:PZT:2001}. New orthorhombic and monoclinic phases have also been found in PZN-$x$PT and 
PMN-$x$PT~\cite{Cox-etal:2001,La-Orauttapong-etal:2002,Noheda-etal:phasePMNPT:2002}. These new phases play 
a key role in explaining the high piezoelectric and electrostrictive responses near the MPB. In PZN-$x$PT, the new phase 
is of orthorhombic symmetry ($Bmm2$) and extends in a narrow concentration range around the MPB ($8\%<x<11\%$) 
with almost vertical phase boundaries on either side (see Fig.~\ref{fig1})~\cite{La-Orauttapong-etal:2002}. This new phase can 
be described as a ``matching'' phase between the rhombohedral and tetragonal phases. This means that the polarization 
vector, instead of being aligned with a particular crystal axis, may be pointing in an arbitrary direction within a plane, 
allowing for a very easy reorientation of the polarization vector~\cite{Fu-Cohen:2000}. Recently, 
a model\cite{Vanderbilt-Cohen:2001} has been proposed, that connects the structural 
features\cite{Noheda-etal:PZT:2001} of the lead relaxor systems with their unusual polarization 
properties\cite{Fu-Cohen:2000}.

In addition to the above structural features, there exists another aspect of the polarization properties of relaxor 
systems, whose precise connection to these structural features has not been established yet. It is now a well recognized 
fact that, as relaxors are cooled from high temperatures, any long range structural change, if it develops, is preceded by 
the appearance and growth of polar nanoregions. The first indirect observation of such regions came from birefringence 
measurements by Burns and Dacol on PMN and PZN~\cite{Burns-Dacol:1983}. These measurements revealed 
deviations from a linear temperature dependence of the birefringence $\Delta n(T)$ (= $n_{\parallel} - n_{\perp}$) at 
a temperature $T_{d}$, {\it i.e.} far above the temperature range in which PZN displays the relaxor behavior. It is 
now believed that this birefringence is due to local asymmetric structural distortions that do not yet have a polar character. 
Since then, several other experimental observations have been reported, that support the appearance of the local 
distortions in lead relaxors below 650~K for PMN and below 750~K for PZN. The most direct ones have come from 
measurements of diffuse X-ray or neutron scattering~\cite{Mathan-etal:1991,You-Zhang:1997}. 
Recently, we have reported diffuse neutron scattering results on a PZN single crystal upon cooling (550~K-295~K), 
which clearly reveal the development of a medium range or mesoscopic order~\cite{La-Orauttapong-etal:2001}. 
The width of the diffuse scattering peak, which is related to a correlation length, is shown to provide a measure of 
the size of the polar nanoregions. At high temperature, the temperature dependence of the correlation length 
is consistent with the Curie-Weiss dependence of the dielectric constant. This indicates the purely dynamic character 
of the polarization, characteristic of paraelectric behavior. At a temperature $T^{\ast } = T_{c}+\delta T$ ($\delta T$ = 40~K) 
in pure PZN, the lifetime of the polar correlations becomes progressively longer, leading to the formation of the polar 
nanoregions. Their formation is accompanied by the development of permanent strain fields, which cause 
the Bragg intensity to increase rapidly (relief of extinction). Other indirect evidence for the formation of the 
polar regions comes from inelastic neutron scattering measurements~\cite{Naberezhnov-etal:1999,Gehring-etal}. 
Particularly noticeable are the results of Gehring \textit{et al.} showing the disappearance of the soft transverse 
optic (TO) mode from the inelastic neutron scattering spectra of PZN and PZN-$x$PT 
(as well as PMN and PMN-$x$PT)~\cite{Gehring-etal}. This disappearance has been attributed to a coupling of the TO and 
TA modes, enhanced by the formation of the polar regions but no precise mechanism has been proposed and the role 
of the polar regions has not been clearly explained. The goal of the present study was 
(i)~to extend the previous PZN study to several reflections in other scattering zones, 
(ii)~to identify more definitely the onset of the polar order and verify its temperature evolution, and 
(iii)~to investigate the local structural and polar order at higher PT concentrations ($x$). 
In order to understand the dynamic behavior of the polar regions, acoustic phonon measurements were also made. 
%
%
%
%

PZN is a prototype relaxor ferroelectric that possesses an ABO$_{3}$ cubic perovskite structure in which the B site 
can be occupied by two different cations, $\frac{1}{3}$Zn$^{2+}$ and $\frac{2}{3}$Nb$^{5+}$. Because of the different 
radii and valences of these two cations, PZN exhibits short-range chemical ordering and local fluctuations 
in composition on a nanometer scale~\cite{Yokomizo-etal:1970,Randall-Bhalla:1990}. It begins to 
transform from a cubic to a rhombohedral phase at $T_{c} \sim 410$~K, which falls in the temperature region of 
the maximum of the dielectric peak. PbTiO$_{3}$(PT), on the other hand, is a normal perovskite-type ferroelectric 
that develops a typical long-range ferroelectric order, when transforming from a cubic to a tetragonal 
phase at $T_{c} \sim 760$~K. In the 1980$^{\prime}$s, a series of complete solid solutions of the relaxor 
PZN and ferroelectric PT were synthesized in the range 0 $\leq x \leq$ 0.2. This led to a very important 
development in the field of ferroelectrics~\cite{Kuwata-etal:1981}. In the present paper, we report the results of 
a diffuse neutron scattering study of several PZN-$x$PT single crystals with $x $=0, 4.5, and 9\% and propose 
a model that provides a physical basis for the easy reorientation of the polarization in lead relaxors.

\begin{table*}
\caption{Samples and experimental parameters.}
\label{experiment}
\begin{ruledtabular}
\begin{tabular}{lllllll}
                         & Mass &                          & Collimation                      & $\lambda_i$& $\hbar \omega$-range  		&$T$-range \\
Sample          &(g)       &Spectrometer&  ($'$ = minutes)              &  (~\AA)  	        & (meV)               	    		& (K)               \\ \hline
PZN                & 0.55    &HB-1                 & 48$'$-20$'$-20$'$-70$'$&  2.45              & 0                   		   	&550-295       \\ 
                        & 	         &HB-1A              & 40$'$-20$'$-20$'$-68$'$ &  2.36              & 0                   		    	& 550-295     \\ 
PZN-4.5\%PT& 2.87   &BT2\footnotemark[1]&60$'$-20$'$-20$'$-open  & 2.36 (1.71)\footnotemark[2]     &0     	&720-300       \\
                         &           &BT7         	        & 20$'$-20$'$-20$'$-open  &  2.47             & 0                   		    	&650-375       \\
                         &           &BT9         	        & 40$'$-22$'$-20$'$-80$'$(or open)  &  2.36 (1.64)\footnotemark[3]&0    	&650-300       \\
                        &            &            	        & 40$'$-44$'$-60$'$-80$'$   &  2.36\footnotemark[4]	    &0-10 	&650-375       \\      
PZN-9\%PT   & 5.15  &HB-1A               & 40$'$-20$'$-20$'$-68$'$   &  2.36            & 0                    	                   	&600-295       \\
                        &           &HB-1        	       & 48$'$-40$'$-40$'$-240$'$  &  2.45\footnotemark[4]          	          &0-14 	&600-400       \\ 
\end{tabular}
\end{ruledtabular}
\footnotetext[1]{Only BT2 was set to measure in the [100]-[010] scattering zone.}
\footnotetext[2]{1.71 $\AA$ was used only for (300) and (310) reflections.}
\footnotetext[3]{1.64 $\AA$ was used only for (400) reflection.}
\footnotetext[4]{Inelastic scan by holding the momentum transfer 
$\vec{Q}$ = $\vec{k}_i$ - $\vec{k}_f$ constant, while scanning the energy transfer 
$\hbar\omega$ = $E_i - E_f$.}
\end{table*}

\section{Experiment}

Single crystals of PZN and PZN-9\%PT were grown by spontaneous nucleation from high temperature solutions, 
using an optimized flux composition of PbO and B$_{2}$O$_{3}$~\cite{Zhang-etal:2000}. The PZN-4.5\%PT 
single crystal was grown by the top-cooling solution growth technique, using PbO flux~\cite{Chen-Ye:2001}. 
All as-grown crystals used in the experiment exhibited a light yellow color and high optical quality.

The neutron experiments were carried out on BT2, BT7, and BT9 triple-axis spectrometers at the NIST Center for 
Neutron Research (NCNR) and on HB1 and HB1A triple-axis spectrometers at the High Flux Isotope Reactor (HFIR) 
of Oak Ridge National Laboratory. Highly oriented pyrolytic graphite (002) (HOPG) was used to monochromate 
and analyze the incident and scattered neutron beams. A HOPG filter was used to suppress harmonic contamination. 
The samples were mounted on an aluminum sample holder, wrapped in a copper foil and held in place with either 
an aluminum or a copper wire. To prevent contamination of the spectra by scattering from aluminum, the sample 
holder was painted with gadolinium oxide or boron nitride paste. The sample was then placed inside a closed-cycle 
helium refrigerator capable of reaching temperatures up to 675~K. It is very important to note here that, if the aluminum 
sample holder is present in the incident neutron beam, it will contribute an extraneous signal near the (200) reflection. 
This signal is generated from the Al (200) peak and overlaps with that of the sample\cite{Koo-etal:2002}, since both have similar 
lattice constants ($\sim$ 4.05 \AA ). The measurements were made around several reciprocal lattice points in 
the [100]-[011] or [100]-[010] scattering planes, which allowed access to the [100], [011], and [111] symmetry directions. 
Data were collected upon cooling and no external electric field was applied. All samples and experimental 
conditions, including the spectrometers, the collimations, the neutron wavelengths ($\lambda $), the range of 
energies ($\hbar \omega $), and the temperature ranges ($T$) are listed in Table~\ref{experiment}.

\section{Bragg and Diffuse scattering in PZN-$x$PT}

Our previous report\cite{La-Orauttapong-etal:2001} contained results on neutron elastic diffuse scattering 
from a PZN single crystal around the (011) Bragg reflection. These results provided a measure of the correlation 
length or size of the polar nanoregions as a function of temperature. In the present work, we have extended the 
investigation of pure PZN to solid solution crystals of PZN-$x$PT with $x$=4.5\% and 9\%, and explored several 
other reflections.

%
%
\begin{figure}[tbp]
\includegraphics[width=1.0\linewidth]{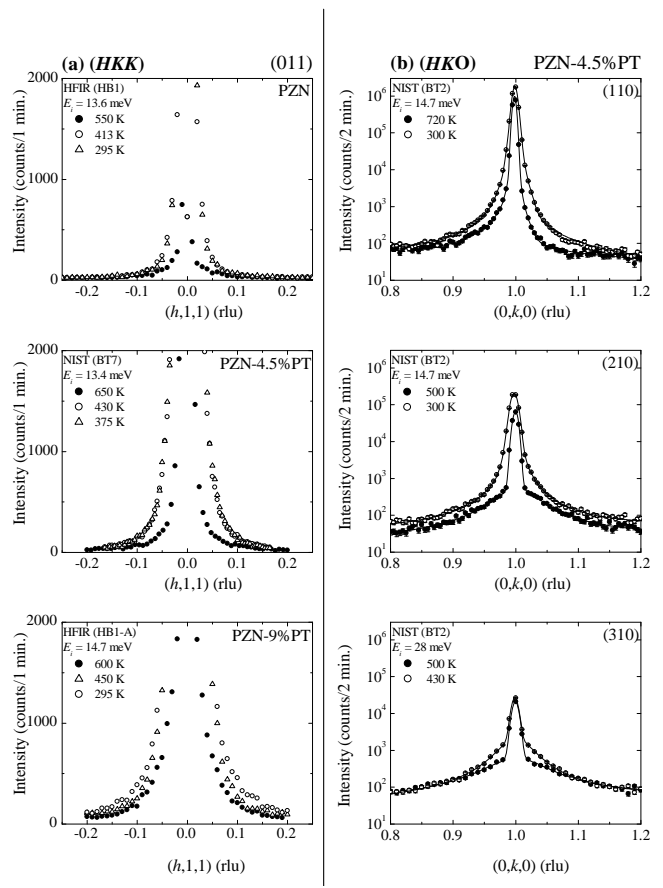}
\caption{Neutron elastic diffuse scattering for PZN-$x$PT (a) in the [100]-[011] scattering zone 
for $x$=0\% (top), $x$=4.5\% (middle), and $x$=9\% (bottom) near the (011) reciprocal lattice point along the 
transverse [100] direction at different temperatures, showing the narrow Bragg peak and the relatively 
broad diffuse scattering peak and (b) in the [100]-[010] scattering zone for $x$=4.5\% at several reflections 
along the transverse [010] direction, showing the relatively low Bragg intensity at higher order reflections. 
Solid lines are fits of the data to a combination of the Gaussian and Lorentzian functions.}
\label{fig2}
\end{figure}
%
%

Typical neutron elastic diffuse scattering spectra around the (011) reflection along [100] direction 
in a [100]-[011] zone are shown in Fig.~\ref{fig2} (a) (left) for $x$=0\%, 4.5\%, and 9\%. Each spectra exhibits a narrow 
Bragg peak and a relatively broad diffuse scattering peak. When the temperature decreases, the diffuse scattering 
peaks broaden and become more extended in the transverse [100] direction in pure PZN and in the diagonal [111] 
direction in PT-doped PZN. The width of the peak also depends on concentration: the peak is broader for the 4.5\%PT 
than for the 9\%PT crystal. The broader peak means a shorter correlation length. Accordingly, in pure PZN 
the correlation length is found to be longer in the [111] direction but with adding PT it becomes longer in 
the [100] direction. It is also longer for 9\%PT than for 4.5\%PT. Below the transition, the width of the diffuse 
scattering peak remains constant, despite a continuing increase in strength of the Bragg peak with decreasing 
temperature. The fitted spectra around several reflection along [010] direction in a [100]-[010] zone are shown 
in Fig.~\ref{fig2} (b) (right) for $x$=4.5\%. This figure shows that, the higher the order reflections, the lower the Bragg 
intensities relative to the diffuse scattering. We will discuss this point later.

%
%
\begin{figure}[tbp]
\includegraphics[width=0.85\linewidth]{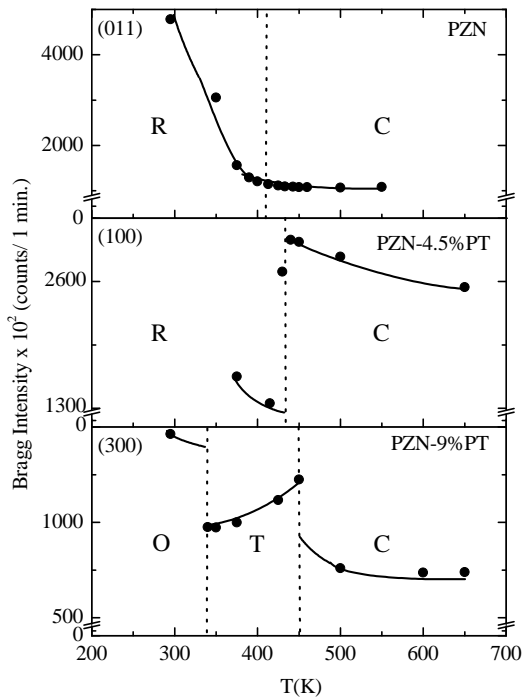}
\caption{Bragg intensities vs. temperature for PZN at (011), PZN-4.5\%PT at (100), and PZN-9\%PT at (300) along 
the diagonal [111] directions, showing the phase transitions. Solid lines are drawn through the data points as guides 
to the eye. Dashed lines mark the phase transitions.}
\label{fig3}
\end{figure}
%
%

The temperature dependence of the Bragg intensities, presented in Fig.~\ref{fig3} for PZN, 4.5\%PT, and 9\%PT, 
reveals the phase transition temperature(s), $T_{c}$. Pure PZN and PZN-$x$PT are seen to behave very differently. 
The former undergoes a continuous structural transition, or freezing, while the latter undergoes abrupt structural 
changes. The particularly strong increase in Bragg intensity, observed for PZN below $T_{c} \sim$ 410~K corresponds 
to the relief of extinction caused by a rapid increase in mosaicity~\cite{La-Orauttapong-etal:2001}. In 4.5\%PT, the Bragg 
intensity increases moderately as the transition is approached but abruptly drops at $T_{c}$ $\sim$ 430~K, 
when the crystal structure transforms from a cubic to rhombohedral symmetry. In the 9\%PT crystal, two sharp 
transitions are observed, respectively at 450~K and at 340~K both marked by an abrupt increase in Bragg intensity. 
From these results, it is evident that the addition of PT triggers sharp structural changes that are otherwise absent. 
The transition temperatures observed in the present work are in agreement with those previously 
reported~\cite{Kuwata-etal:1981} and plotted in Fig.~\ref{fig1}.

\begin{table}[tbp]
\caption{Bragg and diffuse scattering intensities (arbitrary units) in the [100]-[010] scattering zone along the transverse
direction at 500 K in PZN-4.5\%PT crystals.}
\label{diffuse}
\begin{ruledtabular}
\begin{tabular}{lrrc}
$(h k l)$\footnotemark[1]&$I_{Bragg}$&$I_{diffuse}$	& $I_{diffuse}/I_{Bragg}$\\ \hline
1 0 0      &263	           &    5.8	 	&   0.02		\\ 
0 1 1\footnotemark[2]&  982   &  44  	&   0.04		\\ 
2 1 0      &  63	           &   3.3     		&   0.05		\\
3 0 0      &  17	           &   3.9		&   0.23	               \\
3 1 0      &  21	           &   5.6  	 	&   0.27		\\
\end{tabular}
\end{ruledtabular}
\footnotetext[1]{Other observed reflections ({\it i.e.} unmixed ($hkl$) indices, (111), (200), (022), and (400)) have no diffuse scattering.}
\footnotetext[2]{Results from the [100]-[011] scattering zone.}
\end{table}
In order to obtain information about the size and orientation of the polar regions, we fitted the spectra with 
a combination of a Gaussian (Bragg) function and a Lorentzian (diffuse) function~\cite{La-Orauttapong-etal:2001}, 
which was then convoluted with the experimental resolution function, shown as solid lines in Fig.~\ref{fig2} (b). 
The fit was excellent in almost all cases, with $\chi ^{2}$ being close to 1. For higher order reflections, 
(\textit{i.e.} (300) and (310)), the Bragg intensity is relatively low, which makes the data analysis easier. However, 
for lower order reflections the data was difficult to analyze because of the very intense Bragg intensity, which 
was several orders of magnitude stronger than the diffuse intensity part (see Figs.~\ref{fig2} (b) and~\ref{fig3}). 
The fit was nevertheless successful, but the data were fit by the sum of several Gaussian functions plus a 
Lorentzian function. Table~\ref{diffuse} lists the Bragg and diffuse intensities and their ratio ($I_{diffuse}/I_{Bragg}$). 
This ratio is very large at (310) but small at (100). The values found in the present study are in agreement with those 
reported in earlier neutron scattering studies by Mathan \textit{et al.}~\cite{Mathan-etal:1991} and 
Vakhrushev \textit{et al}~\cite{Vakhrushev-etal:1995}. The model calculation of both intensities by 
Mathan \textit{et al.} is close to the polar structure of BaTiO$_{3}$ with antiparallel displacements of cations against 
oxygen ions. 

%
%
\begin{figure}[tbp]
\includegraphics[width=.95\linewidth]{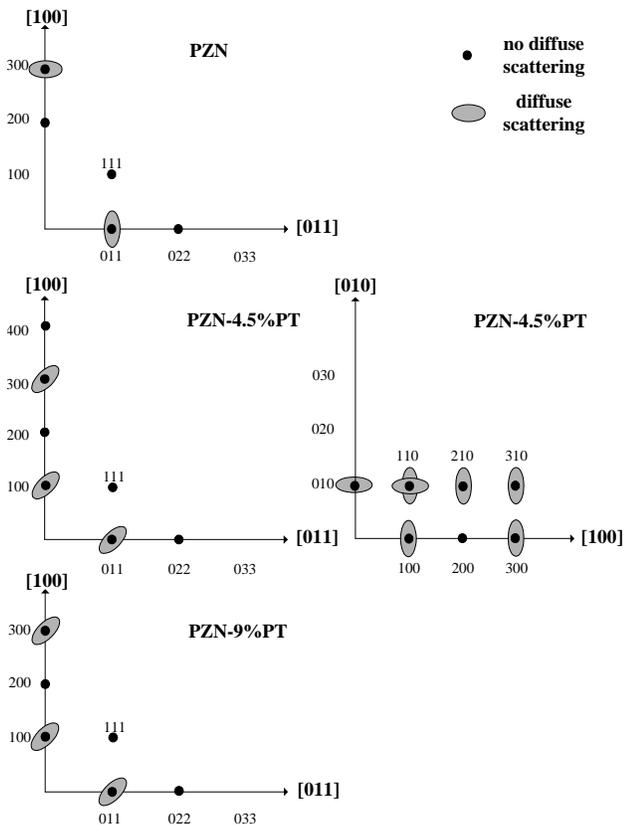}
\caption{Distribution of the diffuse scattering in the scattering plane ([100]-[010] zones) for PZN-$x$PT, 
$x$=0\% (top), $x$=4.5\% (middle), and $x$=9\% (bottom) in the [100]-[011].}
\label{fig4}
\end{figure}
%
%

The diffuse scattering results are summarized in Fig.~\ref{fig4}. Two major observations can be made in this figure. 
First, diffuse scattering is observed near the same reciprocal lattice points for all three concentrations. 
Diffuse scattering was only found for reflections with mixed $hkl$ indices (\textit{i.e.} $oee$ or $eoo$). 
None of the unmixed $hkl$ indices (\textit{i.e.} $eee$ or $ooo$) showed any diffuse scattering (as summarized 
in Table~\ref{diffuse}). Secondly, the diffuse scattering spots or ellipses are preferentially oriented in transverse 
directions for PZN and in the diagonal directions for PZN-$x$PT. No diffuse scattering was found in purely 
longitudinal scans. The first observation contains information on the internal structure of the polar nanoregions. 
Because diffuse scattering can be regarded as the Bragg scattering of a nascent phase, the diffuse intensity 
at different reciprocal lattice points is expected to be proportional to the static structure factors for this new 
phase at these different points. Hence, the fact that diffuse scattering is observed at the same reciprocal 
lattice points for all three concentrations indicates that the internal structure of the new phase, and therefore 
of the polar regions, is the same for all three concentrations. The second observation contains information 
on the preferred directions in which the polar correlations develop. The width of the diffuse scattering being 
inversely proportional to the correlation length ($\Delta q = 2/\xi)$, the change in orientation of the diffuse 
scattering ellipses from PZN to PZN-$x$PT indicates that, upon adding PT, the correlations tend to grow in 
a different direction.

\begin{figure}[tbp]
\includegraphics[width=0.65\linewidth]{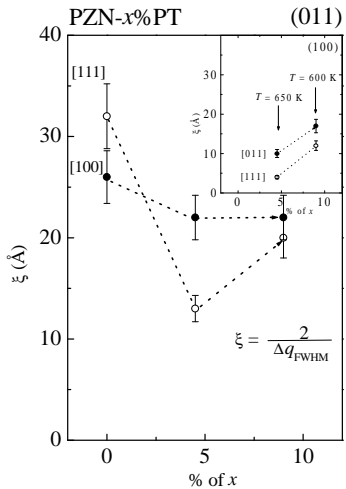}
\caption{$\protect\xi $ vs. concentration $x$ for $x$=0 (at 550~K), $x$=4.5 (at 600~K), and $x$=9 (at 600~K) at the (011) 
point along the [100] and [111] directions in the [100]-[011] zone. An inset is for $x$=4.5 (at 650 K) and $x$=9 (at 600 K) 
at the (100) point along the [011] and [111] directions in the [100]-[011] zone.}
\label{fig5}
\end{figure}

In PZN, the polar regions are preferentially more extended in the diagonal [111] than in the transverse 
[100] or [110] directions, which is consistent with the larger Pb displacement along [111]. However, upon adding PT, 
the preferred orientation evolves to the transverse directions. This trend is clearly illlustrated in Fig.~\ref{fig5}, 
in which the correlation length around the (011) and (100) reflections is shown as a function of concentration 
$x$ and at temperatures far above $T_{c}$. The crossover from one orientation to the other appears to be 
around $x=2\%$. Values of the correlation lengths obtained from measurements around different reflections
in 4.5\%PT at three temperatures, 720~K ($T \leq T_{d}$), 500~K ($T \sim T^{\ast }$), and 300~K ($T < T^{\ast }$), 
are also summarized in Table~\ref{correlation}. For this concentration, the correlation lengths are, respectively, 
about 4-18~\AA~(or 1-4 unit cells), 13-30~\AA~(3-7), and 51-91~\AA~(13-22), small at high temperature and growing 
upon cooling. The longest correlation length is observed around (110) and the shortest one around (310). 
The correlation lengths measured around different reflections are ordered similarly at all three 
temperatures~(\textit{eg.} the size measured at (110) is always found to be larger than the one at (210)). 

The absence of diffuse scattering around the (200) reflection ($hkl$ unmixed) in PZN as well as in PZN-$x$PT 
is of particular significance. A calculation of the dynamic structure factor for a perovskite ferroelectric indicates that 
the diffuse scattering intensity should be large around the (200) reflection, if it were due to correlations associated 
with the usual soft polar mode~\cite{Harada-etal:1970}. The present results therefore indicate that the diffuse scattering 
observed is not due to the soft mode but, rather, to the polar regions that are ubiquitous in relaxor ferroelectrics. 
This important point is discussed in detail in section~\ref{discussion}. 
\begin{table}[tbp]
\caption{Correlation length ($\xi$) around the (100), (011), (210), (300), and (310) reflections in the [100]-[010] 
scattering zone of PZN-4.5\%PT crystals at three temperatures: 720~K ($T \leq T_d$), 500~K ($T \sim T^*$), and 300~K ($T < T^*$).}
\label{correlation}
\begin{ruledtabular}
\begin{tabular}{clcccc}
	&\multicolumn{5}{c}{Correlation length (\AA)} \\
$T (K)$	&\multicolumn{5}{c}{around $(hkl)$ reflections} \\ \hline
	&100\footnotemark[1]	&110	&210	&300	&310	\\ \hline
720\footnotemark[2]	&  10			&  18	&    4	&  10	&     -	\\ \hline
500\footnotemark[3]	& 25(30,18)		&    -	&  15	&  27	&   13	\\ \hline	
300\footnotemark[4]	&  87			&  91	&  51	&  65	&      -	\\
\end{tabular}
\end{ruledtabular}
\footnotetext[1]{Results from the [100]-[011] scattering zone are in parenthesis for the [011] and [111] directions, respectively.}
\footnotetext[2]{At 720 K, $\xi$ $\sim$ 1-4 unit cells.}
\footnotetext[3]{At 500 K, $\xi$ $\sim$ 3-7 unit cells.}
\footnotetext[4]{At 300 K, $\xi$ $\sim$ 13-22 unit cells.}
\end{table}

%
%
\begin{figure}[tbp]
\includegraphics[width=0.85\linewidth]{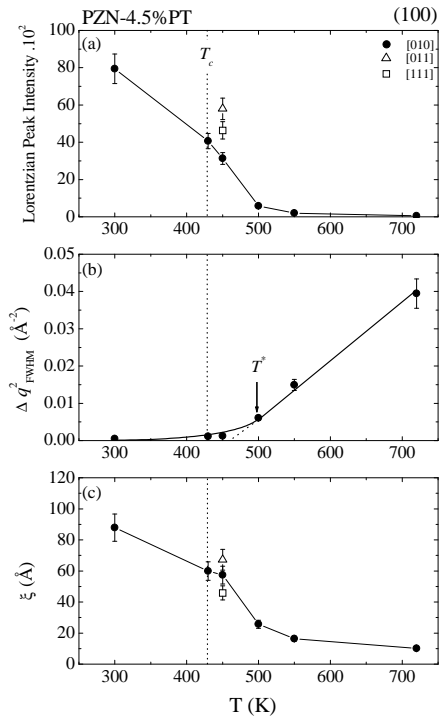}
\caption{Temperature dependence of (a) Lorentzian Peak Intensity, 
(b) $\Delta q_{_{FWHM}}^{2}$, and 
(c) correlation length for PZN-4.5\%PT at the (100) reciprocal lattice point along the [010] direction in the [100]-[010] zone. 
A triangle and a square represent the results (at 450 K) along the [011] and [111] directions in the [100]-[011] zone.}
\label{fig6}
\end{figure}

%
%
The temperature dependencies of the Lorentzian fit parameters, \textit{i.e.} the diffuse intensity (top), 
the square of the diffuse scattering width at half maximum, $\Delta q_{_{FWHM}}^{2}$ (middle), and the correlation 
length (bottom) are presented in Fig.~\ref{fig6} for 4.5\% at the (100) reflection. The diffuse scattering intensity, 
in ferroelectrics, is expected to be proportional to the average polarization squared $\left\langle P_{local}^{2} \right\rangle $. 
In Fig.~\ref{fig6} (a), the diffuse scattering intensity increases continuously with decreasing temperature, first slowly 
down to 500~K, then faster as it approaches $T_{c}=430$ K. Also, the overall diffuse intensity is found to be lower in 
the [111] direction than in the [011] direction. This is consistent with the observations made earlier concerning 
the preferred direction of the correlations in the PT crystals by contrast with the preferred direction in pure PZN. 
In Fig.~\ref{fig6} (b), the full width (at half maximum) squared, $\Delta q_{_{FWHM}}^{2}$, is plotted as a function 
of temperature. As we have explained earlier~\cite{La-Orauttapong-etal:2001}, the initial linear dependence, down 
to 500~K, corresponds to a Curie-Weiss behavior expected in a temperature range in which the polar correlations are 
purely dynamic. At approximately $T^{\ast} \simeq$ 500~K (or $T_{c}+\delta T$, with $\delta T \sim 70$~K), 
$\Delta q_{_{FWHM}}^{2}$ deviates from a straight line and finally levels off at lower temperatures. 
The deviation from this linear dependence marks the appearance of long lived and slowly relaxing polar regions. 
Not unexpectedly, the appearance of the polar regions and the process of slowing down begins at a higher temperature 
in PZN-$x$PT than in pure PZN (for which $\delta T \sim$ 40~K). This fact can be ascribed to the enhanced correlations 
between polar clusters and their stabilization by the tetragonal strain introduced by the presence of PT. 
Below a certain temperature, the size of the polar regions saturates and ``freezing'' may occur, as indicated by 
a constant width of the diffuse scattering peak. In Fig.~\ref{fig6} (c), we present the temperature dependence 
of the correlation length for 4.5\%PT at the (100) reflection along the [010] direction. Upon cooling, the correlation 
length increases continuously, but more rapidly below 500 K, and also tends to saturate at low temperatures. 
It is important to emphasize that the correlation length in this direction is smaller than in the transverse [011] 
direction, as measured in the [100]-[011] scattering plane. This is primarily associated with the short-range 
correlated displacements of the Pb atoms~\cite{La-Orauttapong-etal:2001}.
%
%
%
%

\section{Transverse acoustic phonon (TA) damping}

\begin{figure}[tbp]
\includegraphics[width=0.9\linewidth]{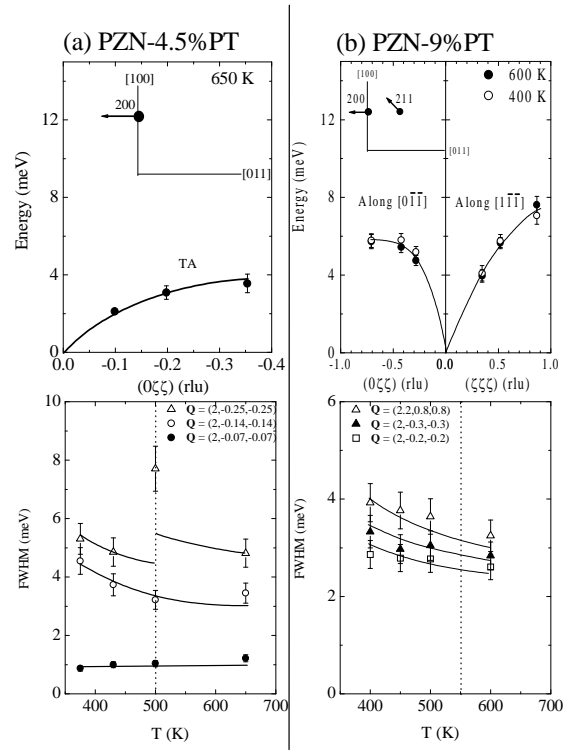}
\caption{Top: (a) dispersion curve of the TA mode in PZN-4.5\%PT at 650 K and 
(b) phonon peak positions taken along [0$\bar{1}$$\bar{1}$] as well as [1$\bar{1}$$\bar{1}$] in PZN-9\%PT at 600 K and 400 K. 
The inset shows scan trajectory. Bottom: (a) temperature dependence of the FWHM of the TA mode in PZN-4.5\%PT 
and (b) temperature dependence of the FWHM of the TA and the [1$\bar{1}$$\bar{1}$] modes in PZN-9\%PT.}
\label{fig7}
\end{figure}

The electrostrictive effect arises from the coupling between polarization and acoustic waves or strain. 
In relaxors, the polarization originates from the polar nanoregions. Because of the unusually strong 
electrostriction exhibited by relaxors~\cite{Park-etal:1997,Viehland-etal:2001}, one should 
expect changes in the behavior of the acoustic phonons when the polar regions appear. Such changes 
should also be expected in coupling between the TO and TA phonons. Recent inelastic neutron scattering 
studies have revealed a very anomalous damping of the TO phonon in lead relaxors~\cite{Gehring-etal}. 
This so-called ``waterfall'' effect has been attributed to the TO-TA coupling. However, because the waterfall 
is not observed in conventional ferroelectrics, it is clear that the TO-TA coupling cannot, by itself, 
explain this effect. The strong electrostriction and the polar nanoregions must, therefore, play an essential role. 
In order to understand this role, we have carried out acoustic phonon measurements in 4.5\%PT and 9\%PT crystals. 
Constant-$\vec{Q}$ scans ($\vec{Q}=\vec{q}+\vec{G}$) were used to collect data on each sample, 
where $q$ is the momentum transfer relative to the $\vec{G}$ = (2,0,0) and (2,1,1) Bragg points, measured along 
the [$0\bar{1}\bar{1}$] and [$1\bar{1}\bar{1}$] symmetry directions, respectively. The peak position of the scattered 
neutron intensity as a function of $q$ (top) and the full width at half maximum (FWHM) as a function of 
temperature (bottom) are presented for 4.5\%PT in Fig.~\ref{fig7} (a) and for 9\%PT in Fig.~\ref{fig7} (b). 
The most important feature of these data is the increase in the FWHM of the TA phonon peak upon cooling 
for intermediate $q$ values. The results obtained on the 4.5\%PT are shown in Fig.~\ref{fig7} (a) 
for $\vec{Q}$ = (2,-0.07,-0.07) (or $q \sim$ 0.10 rlu\cite{rlu}), $\vec{Q}$ = (2,-0.14,-0.14) ($q \sim$ 0.20 rlu), and 
$\vec{Q}$ = (2,-0.25,-0.25) ($q \sim$ 0.35 rlu). For small $q \sim$ 0.10 rlu, the phonon peak remains narrow at 
all temperatures, and may even be narrower at 375~K. This is consistent with the fact that no corresponding 
diffuse scattering was observed at $\vec{Q}$ = (2,0,0). Broadening of the phonon occurs only for longer 
$q^{\prime }s$. For $q \sim$ 0.35 rlu, the peak is already quite broad at 650~K, suggesting that, the greater 
the $q$, the higher the temperature at which the phonon becomes heavily damped. This behavior is definitely 
consistent with the formation of the polar nanoregions, smaller at high temperature and growing with 
decreasing temperature. These results are also consistent with other neutron studies of similar systems. 
For example, in an earlier inelastic neutron study of PMN, Naberezhnov \textit{et al.} reported the broadening 
of the TA mode at $\vec{Q}$ = (2,2,0.2), near the Burns temperature, $T_{d} \sim$ 650~K~\cite{Naberezhnov-etal:1999}. 
Very recently, in a study of PMN-20\%PT, Koo \textit{et al.}\cite{Koo-etal:2002} have also reported a similar increase of 
the TA phonon damping. Finally, it is worth noting that the values of $q$ at which the TA phonon is observed 
to broaden in both the previous studies as well as in the present one, are close to $q_{wf}$ at which 
the waterfall effect has been reported~\cite{Gehring-etal}. 

%
%
%
%
The phonon results obtained on the 9\%PT crystal are shown in Fig.~\ref{fig7} (b). The peak position\cite{rlu} is 
clearly temperature independent in both directions (compare solid and open circles). The width or damping of 
the phonon is greater in the [$1\bar{1}\bar{1}$] direction than in the [$0\bar{1}\bar{1}$] direction. 
However, in the [$0\bar{1}\bar{1}$] direction, it is smaller for this crystal than for the 4.5\%PT one and it does not 
increase as fast with decreasing temperature (compare the FWHM vs. $T$ at $\vec{Q}$ = (2,-0.2,-0.2) or $q \sim$ 0.28 rlu). 
This observation suggests that the TA damping decreases with increasing PT concentration. So far 9\%PT is 
the highest concentration that we have studied. It would be interesting to follow the evolution of the TA damping 
in PZN-$x$PT crystals with higher $x$.

\section{Discussion}
\label{discussion}

It is by now well established that compositional and/or site disorder brings about the formation of local polar 
nanoregions in mixed perovskite ferroelectrics, including PMN, PZT, and PZN. These regions result from 
short-range correlated ionic displacements~\cite{Mathan-etal:1991}. They nucleate at a temperature 
well above the transition temperature, $T_{c}$, and are dispersed as islands throughout the lattice. 
Diffuse neutron and X-ray scattering provide the most direct evidence for the morphology of these regions.

\begin{figure}[tbp]
\includegraphics[width=.85\linewidth]{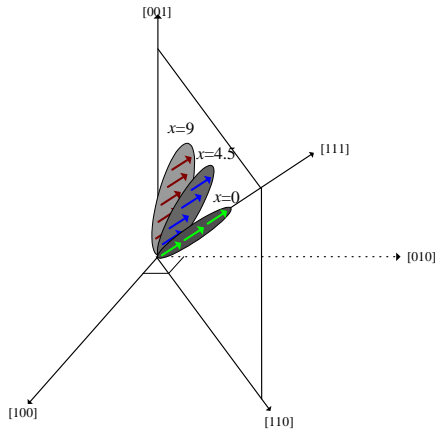}
\caption{Polarization vector and growth direction of the polar nanoregions in PZN-$x$PT with increasing concentration $x$.}
\label{fig8}
\end{figure}

Diffuse scattering results provide two distinct types of information. The reciprocal lattice points around which 
diffuse scattering is observed provide information on the symmetry of the local lattice distortions or internal 
structure of the polar regions, and the shape or distribution of the diffuse scattering intensity around these 
points gives information on the morphology of the polar regions or the direction in which they grow. The \textit{first} 
type of information concerns the direction of the polarization vector within the polar regions, and the \textit{second} 
type, the orientation of the strain fields associated with them. In PZN-$x$PT, the present data show that elastic 
diffuse scattering is only observed around points where the ($hkl$) indices have mixed parities, in particular 
(100), (110), (210), (300), and (310). Diffuse scattering is also strongest around (110). These results are identical 
to those reported for pure PZN, which we interpreted earlier as revealing the rhombohedral internal structure 
of the polar regions, consistent with the [111] displacement of the Pb$^{2+}$ ions~\cite{La-Orauttapong-etal:2001}. 
If the internal structure and the direction of the polarization vector are the same in both PZN and PZN-$x$PT, 
the direction in which the polar regions grow is different in both systems. In pure PZN, the diffuse scattering 
was found to be more extended in a transverse direction ([100] or [011]), indicating the growth of the polar regions 
in a [111] direction. By contrast, in PZN-$x$PT, the diffuse intensity becomes more extended in the diagonal [111], 
indicating the preferential growth of the polar regions in a [100] or [011] transverse direction. As shown in 
Fig.~\ref{fig5}, the higher the PT concentration and the closer to a tetragonal [100], the higher the ratio of 
the correlation lengths in the transverse and [111] directions. This trend is represented schematically 
in Fig.~\ref{fig8}, for PZN, 4.5\%PT, and 9\%PT in a [110]-[001] plane. Thus, with increasing PT, the local 
polarization direction remains unchanged ([111] direction), but the polar regions now preferentially grow 
in a different direction, closer to tetragonal.
%
%
%
%

The picture presented in Fig.~\ref{fig8} illustrates the concept of local polarization and strain fields pointing in two 
different directions. The present picture provides a physical basis for the model proposed earlier by Fu and Cohen, 
in which the macroscopic polarization vector may be pointing in an arbitrary direction within 
a plane~\cite{Fu-Cohen:2000}. The ease of rotation of the macroscopic polarization can now be 
understood to be due to the fact that the local strain fields only need to reorient by a much smaller angle than 
the local polarization. However, it is important to note that, although the results reported here clearly support 
such a picture, the diffuse scattering ellipse observed in a particular scattering plane may only be the projection 
of the 3D scattering ellipsoid. Additional measurements would be necessary to map out entirely the diffuse 
scattering ellipsoids. The temperature dependence of the diffuse scattering intensity in PZN and PZN-$x$PT 
further demonstrates that it is related to the appearance of the polar nanoregions. Another equally strong element of 
proof is the absence of diffuse scattering around the (200) reflection. This was checked in all possible directions 
and in two different scattering zones [100]-[011] and [100]-[010]. Because, in perovskites, the static structure 
factor that corresponds to the soft phonon mode is largest at the (200) reflection, the absence of diffuse 
scattering around this point clearly indicates that it is related to the presence of the local order that develops 
below 700-600 K. A similar result had been obtained earlier by one of us in two different relaxors, 
K$_{1-x}$L$_x$TaO$_3$ (KLT) and KTa$_{1-x}$Nb$_x$O$_3$ (KTN) ~\cite{Yong-etal:2000}. 
Hirota \textit{et al.}\cite{Hirota-etal:2002} have recently reported a similar result in PMN and have proposed the concept of 
the phase-shifted condensed soft mode, to account for the contribution of the polar nanoregions. In their dynamic 
calculation, the atomic displacements include both \textit{the optic mode displacement} $(\delta _{cm})$, 
satisfying a center of mass condition and \textit{the single phase shift} $(\delta _{shift})$ within polar nanoregions.

The result presented above also reveal three important temperature ranges in the behavior of PZN and PZN-$x$PT, 
which seems to be characteristic of all relaxor ferroelectrics. (i) At high temperature, $T^{\ast }<T<T_{d}$, the correlation 
length squared, $\xi ^{2}$, is inversely proportional to temperature. Because $\xi ^{2}$ is also proportional to 
the dielectric constant, this temperature dependence corresponds to a Curie-Weiss law, which indicates that, 
in this range, the polarization is entirely dynamic and the system behaves as a normal paraelectric. 
(ii) Below $T^{\ast }$, the deviation from this Curie-Weiss law reflects the appearance of long-lived polar fluctuations 
in the crystal, accompanied by local strain fields; it is also in this range that the dielectric constant begins to exhibit 
a strong frequency dispersion, the relaxor behavior. This behavior is clearly due to the reorientation of 
the polarization or, as seen from the structural results reported here, of the long-lived polar nanoregions. 
With the addition of PT, $T^{\ast }$ increases from $T_{c}$+40 K in PZN to $T_{c}$+70 K in 4.5\%PT. This trend is 
due to enhanced correlations between polar regions with increasing PT concentration. (iii) As the temperature is 
further decreased below $T^{\ast }$, and as the polar regions grow, their local strain fields increase in strength 
and their reorientation becomes slower. In PZN, it is believed that there is no overall preferred [111] orientation, 
so that individual polar nanoregions eventually freeze out with their polarization in one of eight possible [111] 
axial directions, each according to their local strain field. Hence, the mosaicity increases and, due to the relief 
of extinction below $T_{c}\sim 410$ K, the Bragg intensity rapidly rises. This interpretation is borne out by the fact 
that the Bragg intensity remains high at lower temperatures. It is also interesting to note that, simultaneously, 
the dielectric constant decreases rapidly at lower temperatures, which also confirms the freezing-in of the polar 
regions in PZN.

The addition of PT increases the magnitude of the local distorsions, with a tendency towards tetragonal symmetry, 
as in PbTiO$_{3}$.  In all of this, it is important to remember that strain is the dominant cause of orientational freezing 
and that the strain energy is the same whether the polarization points up or down. Further addition of PT increases 
the tetragonal strain, leading to an abrupt structural transition in 4.5 \%PT at 430 K and two in 9\%PT, at 450 K and 
340 K, respectively. The abrupt character of these transitions suggests the first order character of strain-driven transitions. 

In summary, the present elastic scattering results can be conveniently interpreted within the framework of three main 
temperature regions: 
(i) dynamic polarization fluctuations at high temperature ($T\sim 700$K$-600$K), 
(ii) condensation of polar nanoregions that can still reorient as a unit and progressive slowing down of 
the reorientational motion at intermediate temperatures ($T\lesssim T^{\ast }$), and 
(iii) orientational freezing of the polar nanoregions at low temperatures ($T<T^{\ast }$), with or without an explicit 
structural transition at $T=T_{c}$ .

When investigating the polarization dynamics of relaxors, it is also important to examine the TA phonons, 
since these can couple to the reorientation of the localized strain fields that are known to accompany 
the formation of the polar nanoregions~\cite{Rowe-etal:1979}. In PMN and PMN-20\%PT, Naberezhnov {\it et al.} 
and Koo {\it et al.}\cite{Naberezhnov-etal:1999,Koo-etal:2002} have reported broadening of the TA phonon 
starting at $T_{d}$, and near the wavevector $q_{wf}$ at which the TO phonon has been reported to 
disappear (``waterfall'')~\cite{Gehring-etal}. These authors have linked the onset of damping of the TA and 
the onset of birefringence at $T_{d}$ to the appearance of the polar nanoregions. However, both these 
results only indicate the appearance of local distortions and strain fields but do not provide information as to 
the possible polar character of these local distortions. In our measurements of 4.5\%PT and 9\%PT, 
damping of the TA phonon is seen to increase, starting at $T^{\ast}$. It also starts at a large $q$ first and 
at smaller $q$ with decreasing temperature. In other words, we find that, at a given temperature, 
the larger the $q$, the higher the damping. In fact, we do expect such a trend from the coupling to smaller 
distorted regions at higher temperatures and to larger and slower ones at lower temperatures. 
In fact, our results show that, in the temperature range $T^{\ast }<T<T_{d}$, the correlation length follows 
a Curie-Weiss law, characteristic of a paraelectric state in which the polarization is completely dynamic. 
The polar nanoregions only appear in earnest at $T^{\ast }$, which is also the temperature at which the dielectric 
constant begins to exhibit frequency dispersion, {\it i.e.} the relaxor behavior. As seen from Fig.~\ref{fig7} (a) 
for 4.5\%PT, the phonon damping begins to increase at $\sim $ 500 K (\textit{i.e.} at $T^{\ast }$) 
and $\vec{Q}$ = (2,-0.14,-0.14) (or $q \sim$ 0.20 rlu), which corresponds to about 5 unit 
cells~\cite{correlation at 500K}. This result is consistent with the size of the polar regions derived from our 
diffuse neutron elastic scattering data (Table~\ref{correlation}). Such an agreement provides evidence 
that the increase in TA phonon damping is connected to the appearance of the polar regions. The hypersonic 
damping properties studied by the Brillouin scattering has provided a clear indication of ferroelectric ordering 
in complex relaxor ferroelectric~\cite{Tu-etal:1995}. In 9\%PT, the TA phonon damping is lower than in 
4.5\%PT (bottom of Fig.~\ref{fig7}) but shows a similar trend. It is important to note that the measured TA phonon 
corresponds to the C$_{44}$ elastic modulus, which can couple to the reorientations of a strain field 
with rhombohedral symmetry between different [111] directions. The higher frequency of the TA phonon 
in 9\%PT than in 4.5\%PT (top of Fig.~\ref{fig7}) indicates that C$_{44}$ is higher in 9\%PT. Both observations, 
lower damping and higher TA phonon frequency, suggest that, with increasing PT, the polar regions are less 
able to reorient and the lattice becomes more rigid. Further work is in progress on the TA phonons as well as 
on the low frequency rotational dynamics associated with the reorientation of the polar regions in single crystals 
of PZN-$x$PT.

\section{conclusions}

Neutron \textit{elastic} diffuse scattering in PZN, PZN-4.5\%PT, and PZN-9\%PT appears at temperatures well above 
the phase transition ($\sim$ 700-600 K), marking the formation of local distortions at $T_{d}$, which acquire a polar character 
at the lower temperature $T^{\ast }$. In both pure PZN and PZN-$x$PT, diffuse scattering is observed only around 
reciprocal lattice points with mixed parity indices, which is indicative of the rhombohedral internal symmetry of 
the polar nanoregions. From the orientation of the diffuse scattering ellipsoids around these points, it appears that, in PZN, 
the correlations preferentially develop along a diagonal or [111] direction. With increasing PT, these correlations shift 
towards a transverse direction, most likely [100], consistent with the widening of the tetragonal phase field for higher PT 
concentrations. Nonetheless, it is important to emphasize that, with the addition of PT, the internal structure of the 
polar regions remains rhombohedral. This distinction between the [111] distortion of the local unit cells and 
the [100] orientation of the correlations may constitute an essential feature of the Pb-relaxor ferroelectrics. 
Upon cooling, PZN and PZN-$x$PT undergo three distinct stages: purely dynamic polarization about $\sim$ 700-600 K, 
reorienting polar regions and, finally, polar regions with fixed orientations due to freezing in pure PZN and 
phase transitions in PZN-$x$PT. In the middle range, it appears that the reorienting polar regions couple 
to the TA phonons, leading to their softening and increased damping. Comparison of the phonon results with 
the temperature and $q$ dependence of the diffuse scattering results does indeed suggests that the TA broadening 
originates from the reorientation of strain fields associated with the polar nanoregions. This may be a common feature 
of many relaxor systems.

\begin{acknowledgments}
This work was supported by DOE Grant No. DE-FG02-00ER45842 and ONR Grant No. N00014-99-1-0738 (Z.-G. Ye). 
We acknowledge the support of the NIST Center for Neutron Research as well as the Oak Ridge National Laboratory 
in providing the neutron facilities used in this work. One of the authors (D. La-Orauttapong) was supported by a 
Royal Thai Government scholarship during this work. It is pleasure to acknowledgment 
R.K. Pattnaik, S. A. Prosandeev, and O. Svitelskiy for helpful discussions.
\end{acknowledgments}

\end{document}